\documentclass[twocolumn, pra, showpacs,superscriptaddress]{revtex4}

\usepackage{amssymb}
\usepackage{graphicx}
\usepackage{dcolumn}
\usepackage{bm}
\usepackage{amsmath}

\begin{document}

\title{Density-functional theory for the spin-1 bosons in a one-dimensional
harmonic trap }
\author{Hongmei Wang}
\affiliation{Institute of Theoretical Physics, Shanxi University, Taiyuan 030006, P. R. China}
\affiliation{Department of Physics, Taiyuan Normal University, Taiyuan 030001, P. R. China}
\author{Yunbo Zhang}
\email{ybzhang@sxu.edu.cn}
\affiliation{Institute of Theoretical Physics, Shanxi
University, Taiyuan 030006, P. R. China}

\begin{abstract}
We propose the density-functional theory for one-dimensional
harmonically trapped spin-1 bosons in the ground state with repulsive
density-density interaction and anti-ferromagnetic spin-exchange interaction. The
density distributions of spin singlet paired bosons and polarized bosons
with different total polarization for various interaction parameters are
obtained by solving the Kohn-Sham equations which are derived based on the
local density approximation and the Bethe ansatz exact results for homogeneous 
system. Non-monotonicity of the central densities is 
attributed to the competition between the density interaction and spin-exchange.
The results reveal the phase separation of the 
paired and polarized bosons, the density profiles of which respectively approach the 
Tonks-Girardeau gases of Bose-Bose pairs and scalar bosons in the case of strong
interaction. We give the R-P phase diagram at strong interaction and find the critical 
polarization, which paves the way to direct observe the exotic singlet pairing
in spinor gas experimentally. 
\end{abstract}

\maketitle

\section{Introduction}

Spinor Bose gases and one-dimensional (1D) system are both the
fascinating topics in the research of cold atoms \cite%
{Kawaguchi2012,Kurn2012,Cazalilla2011,GuanRMP}. The studies of their crossing point for
1D spinor bosons are also attractive in theories \cite%
{Wenxian2005PRA71,yajiang2006,Cao2007,Deuretzbacher2008,Girardeau2011,Essler2009,Lee2009,Kuhn2012}
and have been realized in recent experiments \cite%
{Liao2010,Campbell2012,Vinit2012}.
Spinor gases are prepared in the optical traps where the induced electric
dipole moment determines the laser-atom interaction and involve an ensemble
of Bose atoms condensed in a coherent superposition of all possible
hyperfine states. The early experimentally achieved spinor gases include $%
^{23}$Na \cite{Kurn1998,Miesner1999} and $^{87}$Rb 
\cite{Matthews1999,Barrett2001}. In the three-dimensional (3D) case, the
spin-dependent spin-exchange interactions are much weaker than the
spin-independent short-range density-density interactions, for example, the
ratio of them are $c_{2}/c_{0}\left( \text{Na}\right) =0.03$ \cite%
{Crubellier1999} and $c_{2}/c_{0}\left( \text{Rb}\right) =-0.005$ \cite%
{Kempen2002} respectively. The ground-state wavefunction is represented by a
spinor wavefunction which minimizes the free energy \cite%
{Ohmi1998,Tianlun1998,Koashi2000} and the spin-exchange interactions give
rise to a rich variety of phenomena such as spin domains \cite{Miesner1999},
textures \cite{Ohmi1998}, spin mixing dynamics \cite%
{Law1998,Pu1999,Wenxian2005PRA72,zhangjie2011}, and fragmentation of condensate \cite%
{Tianlun2000,Mueller2006,zhangjie2010} etc. On the other hand, 1D systems can be realized
by confining the cold atoms in strong anisotropic traps where the motion
of atoms is effectively 1D \cite%
{Greiner2001,Moritz2003,Paredes2004,Kinoshita2004,Haller2009,Liao2010,
Campbell2012,Vinit2012}. 
The interaction among the atoms can be tuned in the whole regime of
interaction strength via the idea of Feshbach resonance as well as the
confinement-induced resonance (CIR) \cite{Olshanii1998,Bergeman,Sinha}.
These experimental developments have provided unprecedented opportunities
for testing the theory of 1D exactly solvable many-body models \cite%
{Lieb1963,Yang1967,Yang1968,Fuchs2004,Imambekov2006,Imambekov2003,Sutherland1968,Cao2007}.

Many theories have studied the 1D spinor gases. Under the mean-field
theory, Zhang and You checked the validity of a Gaussian ansatz for the
transverse profile in the weak interaction regime and a Thomas-Fermi ansatz
(TFA) in the strong interaction regime \cite{Wenxian2005PRA71}. Hao \textit{%
et al.} \cite{yajiang2006} modified the Gross-Pitaevskii (GP) equations
based on the solution of Lieb-Lininger (L-L) \cite{Lieb1963} model to reveal
that the total densities of the 1D spinor bosons exhibit the Tonks-Girardeau
(TG) \cite{Girardeau1960} properties of 1D scalar bosons when
density-density interaction is strong enough. The detailed TG and super
Tonks-Girardeau gas (STG) properties of 1D spinor bosons have been
investigated particularly by Deuretzbacher \textit{et al.} \cite%
{Deuretzbacher2008} and Girardeau \textit{et al. }\cite{Girardeau2011} with
the method of Bose-Fermi mapping.

If the spin-exchange interaction can be modulated to the order of
density-density interaction, the competition between these two kinds of
interaction must be considered. Cao \textit{et al.} find that the 1D
homogeneous spinor bosons under $c_{0}=c_{2}$ can be exactly solved with
Bethe ansatz (BA) method \cite{Cao2007}. Essler \textit{et al. }show its
low-energy degrees of freedom are equivalent to a spin-charge separation
theory of the U(1) Tomonaga-Luttinger liquid describing the charge sector
and the O(3) nonlinear $\sigma $ model describing the spin sector 
\cite{Essler2009}. By means of the thermodynamical Bethe ansatz (TBA) method
\cite{Lee2009,Kuhn2012}, Lee \textit{et al.} and Kuhn \textit{et al.} give the ground state 
phase diagram and investigate the universal thermodynamics and quantum 
criticality of the trapped 1D spinor bosons for the strong interaction situation.

So far, a method is not available for the 1D trapped spin-1 bosons in the
entire region of interaction from weak to strong. In this paper, we develop
the Hohenberg-Kohn-Sham density functional theory (DFT) to investigate the
ground-state properties of 1D harmonically trapped spin-1 bosons. DFT has
been widely used for treating electron systems with long-range Coulomb
interaction \cite{HohenbergKohn,Dreizler1990}. It also has been successfully
generalized to cold atom systems with short-range contact interaction \cite%
{Nunes,Kim,Albus,Ma2012}. For 1D cold-atom systems, the method of DFT based
on BA results has been developed to solve the ground state problem of bosons
\cite{Kim,Brand,Yajiang2009}, fermions \cite{Astrakharchik,Xianlong} and
Bose-Fermi mixtures \cite{Hongmei2012}.

Here we apply this method to study how the ground state of 1D trapped spin-1
bosons evolves along with the interaction strength from weak to strong. We
derive the Kohn-Sham (KS) equations by combing the BA solutions and Local
Density Approximation (LDA). The ground state densities and energies for
different interactions are obtained by solving these equations iteratively.
The paper is organized as follows. We introduce our theory in Sec. II and
show the numerical results in Sec. III. The theory part includes the model,
the BA equation for homogeneous system and the KS equations for trapped
system. In the results part, we first show the case that all bosons are
fully paired and then for the partially paired case.

\section{Theory}

\subsection{Model}

We consider $N$ spin-1 bosons of mass $m$ confined in extremely anisotropic
crossed optical dipole traps with delta-function type density-density
interaction and spin exchange interaction between atoms. The trap is
characterized by the radial and axial angular frequencies $\omega _{\bot }$
and $\omega $ with corresponding harmonic oscillator lengths $a_{\bot }=%
\sqrt{\hbar /m\omega _{\bot }}$ and $a=\sqrt{\hbar /m\omega }$ respectively.
When $\omega _{\bot }$ $\gg $ $\omega $, the radial Thomas-Fermi radius is
small enough such that only axial spin domains could form \cite%
{Campbell2012,Vinit2012}. In first quantized form, the Hamiltonian for 1D
spin-1 gas can be written as

\begin{eqnarray}
H &=&\sum_{i=1}^{N}\left( -\frac{\hbar ^{2}}{2m}\frac{\partial ^{2}}{%
\partial x_{i}^{2}}+\frac{1}{2}m\omega ^{2}x_{i}^{2}\right)  \notag \\
&&+\sum_{i<j}\left[ c_{0}^{1D}+c_{2}^{1D}\mathbf{S}_{i}\cdot \mathbf{S}_{j}%
\right] \delta \left( x_{i}-x_{j}\right) ,  \label{hamiltonian}
\end{eqnarray}%
where $\mathbf{S}_{i,j}$ are spin-1 operators, $c_{0}^{1D}$ and $c_{2}^{1D}$
are 1D interaction parameters which can be expressed through interaction
parameters $g_{S}^{1D}$ in total spin $S=0,2$ channels as $c_{0}^{1D}=\left(
g_{0}^{1D}+2g_{2}^{1D}\right) /3$ and $c_{2}^{1D}=\left(
g_{2}^{1D}-g_{0}^{1D}\right) /3$. In experiments, $g_{S}^{1D}$ can be tuned
with $a_{\bot }$ and 3D s-wave scattering length $a_{S}^{3D}$ according to $%
g_{S}^{1D}=2\hbar ^{2}a_{S}^{3D}/ma_{\bot }^{2}\left( 1-Aa_{S}^{3D}/a_{\bot
}\right) $ with constant $A=1.0326$ \cite{Olshanii1998}. Thus $c_{0}^{1D}$
and $c_{2}^{1D}$ may be manipulated in wider range comparing with 3D spinor system.

The number of atoms $N_{+}$, $N_{0}$ and $N_{-}$ corresponding to spin states $%
s=+1,0,-1$ are not conserved because the scattering between two atoms of
spin $s=\pm 1$ can produce two atoms of spin $s=0$ and vice versa, whereas
the total number of atoms $N=N_{+}+N_{0}+N_{-}$ and total spin in the $z$
component $S^{z}=N_{+}-N_{-}$ are conserved yet. Therefore we may consider
the system is composed of two parts of atoms, particle I and particle II
with total particle numbers $N=N_{1}+2N_{2}$ and total polarization $%
P=N_{1}/N$. The number of particle I is $N_{1}=S^{z}$ where all atoms have
parallel spin forming the ferromagnetic phase. The number of particle II is $%
2N_{2}$ where $N_{2}$ pairs of atoms are formed between two spin states $%
s=\pm 1$ or between two spin states $s=0$. Here we have supposed that $%
N_{+}\geqslant N_{-}$ and that $N_{0}$ is even. The sign of $c_{0}^{1D}$
determines that the interactions between the bosons are repulsive or
attractive, while the sign of $c_{2}^{1D}$ determines that the spin exchange
interaction in the pairs are ferromagnetic or anti-ferromagnetic.

\subsection{Bethe ansatz equations for homogeneous system}

With the particles confined in a finite 1D tube with length $L$ instead of a 
harmonic trap as in the Hamiltonian (\ref{hamiltonian}), the system has been exactly solved 
by Cao \textit{et al.} \cite{Cao2007} with BA method under a special condition
\begin{equation*}
c_{0}^{1D}=c_{2}^{1D}=g>0,
\end{equation*}%
i.e., the repulsive density-density interaction equals the antiferromagnetic
spin-exchange interaction. They gave the BA equations
\begin{equation}
e^{ik_{j}L}=\prod\limits_{i=1,i\neq j}^{N}\tilde{e}_{4}\left(
k_{j}-k_{i}\right) \prod\limits_{\alpha =1}^{2N_{2}}\tilde{e}_{-2}\left(
k_{j}-\Lambda _{\alpha }\right) ,  \label{BA 1}
\end{equation}%
\begin{equation}
\prod\limits_{i=1}^{N}\tilde{e}_{2}\left( \Lambda _{\alpha }-k_{i}\right)
=-\prod\limits_{\beta =1}^{2N_{2}}\tilde{e}_{2}\left( \Lambda _{\alpha
}-\Lambda _{\beta }\right) ,  \label{BA 2}
\end{equation}%
with $\tilde{e}_{n}\left( x\right) =\left( x+inc^{\prime }\right) /\left(
x-inc^{\prime }\right) $, $c^{\prime }=c/4$ and $c=2mg/\hbar ^{2}$. $\left\{
k_{i}\right\} $ in the equations is the set of quasi-momentum and $\left\{
\Lambda _{\alpha }\right\} $ is the set of the spin rapidity. The ground
state energy of the system is%
\begin{equation}
E=\frac{\hbar ^{2}}{2m}\sum_{i=1}^{N}k_{i}^{2}.  \label{energy}
\end{equation}

For positive $c_{2}^{1D}$, the spin-exchange interaction is
antiferromagnetic. $2N_{2}$ particles form spin singlet bound pairs between
two spin $s=\pm 1$ atoms or between two spin $s=0$ atoms, whereas $N_{1}$
particles are polarized. The equations (\ref{BA 1}) and (\ref{BA 2}) have $%
N_{1}$ real solutions for $k_{i}$ ($i=1,\cdots N_{1}$) and $N_{2}$ pairs
conjugate complex solutions or string solutions for $k_{\alpha }$ and $%
\Lambda _{\alpha }$ ($\alpha =1,\cdots ,2N_{2}$), i.e.%
\begin{eqnarray}
k_{\alpha } &=&\lambda _{l}\pm ic^{\prime },  \notag \\
\Lambda _{\alpha } &=&\lambda _{l}\pm ic^{\prime },
\label{conjugate complex}
\end{eqnarray}%
with $\lambda _{l}$ ($l=1,\cdots ,N_{2}$) real numbers \cite{Cao2007,Lee2009}%
. Inserting (\ref{conjugate complex}) into (\ref{BA 1}) (\ref{BA 2}) and
adopting the thermodynamic limit, i.e., the length of gas $L\rightarrow
\infty $ and particle numbers $N,N_{1},N_{2}\rightarrow \infty $ with
densities $n=N/L$, $n_{1}=N_{1}/L$ and $n_{2}=N_{2}/L$ finite, we easily
arrive at the BA integral equations
\begin{eqnarray}
2\pi \rho _{1}\left( k\right) &=&1+2\int_{-B}^{B}dk^{\prime }\rho _{1}\left(
k^{\prime }\right) a_{4}\left( k,k^{\prime }\right)  \notag \\
&&+2\int_{-Q}^{Q}d\lambda \rho _{2}\left( \lambda \right) \left(
a_{5}-a_{1}\right) \left( k,\lambda \right) ,  \label{iBA 1}
\end{eqnarray}%
\begin{eqnarray}
\pi \rho _{2}\left( \lambda \right) &=&1+\int_{-B}^{B}dk\rho _{1}\left(
k\right) \left( a_{5}-a_{1}\right) \left( k,\lambda \right)  \notag \\
&&+\int_{-Q}^{Q}d\lambda ^{\prime }\rho _{2}\left( \lambda ^{\prime }\right)
\left( a_{6}+a_{4}-a_{2}\right) \left( \lambda ,\lambda ^{\prime }\right) ,
\label{iBA 2}
\end{eqnarray}%
where
\begin{equation}
\left( a_{n}-a_{m}\right) \left( x_{1},x_{2}\right) =a_{n}\left(
x_{1},x_{2}\right) -a_{m}\left( x_{1},x_{2}\right) ,  \label{an-am}
\end{equation}%
and%
\begin{equation}
a_{n}\left( x_{1},x_{2}\right) =\frac{n\left\vert c^{\prime }\right\vert }{%
\left( nc^{\prime }\right) ^{2}+\left( x_{1}-x_{2}\right) ^{2}},  \label{an}
\end{equation}%
with $\rho _{1}\left( k\right) $ and $\rho _{2}\left( \lambda \right) $
densities of $k$ and $\lambda $. The integral boundaries $B$ and $Q$ are
determined by the conditions%
\begin{eqnarray}
n_{1} &=&\int_{-B}^{B}dk\rho _{1}\left( k\right) ,  \label{n1} \\
n_{2} &=&\int_{-Q}^{Q}d\lambda \rho _{2}\left( \lambda \right) .  \label{n2}
\end{eqnarray}%
From (\ref{energy}), we get the ground state energy per unit length
\begin{eqnarray}
\frac{E}{L} &=&\frac{\hbar ^{2}}{2m}\int_{-B}^{B}dkk^{2}\rho _{1}\left(
k\right)  \notag \\
&&+\frac{\hbar ^{2}}{2m}\int_{-Q}^{Q}d\lambda 2\lambda ^{2}\rho _{2}\left(
\lambda \right) -\epsilon _{b}n_{2},  \label{i energy}
\end{eqnarray}%
where the binding energy of the pair in defined as
\begin{equation}
\epsilon _{b}=\frac{\hbar ^{2}}{2m}\frac{c^{2}}{8}.  \label{binding energy}
\end{equation}

To solve the BA equations, we introduce the dimensionless L-L
interaction parameter $\gamma =c/2n=mg/\hbar ^{2}n$ and polarization
parameter $p=n_{1}/n$. Let $k=Bx$, $\lambda =Qy$, $B=c/\beta _{1}$ and $%
Q=c/\beta _{2}$, the densities of quasi-momentum and spin rapidity turn out
to be $\rho _{1}\left( k\right) =g_{1}\left( x\right) $ and $\rho _{2}\left(
\lambda \right) =g_{2}\left( y\right) $. The integral BA equations (\ref{iBA
1}) and (\ref{iBA 2}) are translated into%
\begin{eqnarray}
2\pi g_{1}\left( x\right) &=&1+\frac{2}{\beta _{1}}\int_{-1}^{1}dx^{\prime
}g_{1}\left( x^{\prime }\right) b_{4}\left( \frac{x}{\beta _{1}},\frac{%
x^{\prime }}{\beta _{1}}\right)  \label{g1} \\
&&+\frac{2}{\beta _{2}}\int_{-1}^{1}dyg_{2}\left( y\right) \left(
b_{5}-b_{1}\right) \left( \frac{x}{\beta _{1}},\frac{y}{\beta _{2}}\right) ,
\notag
\end{eqnarray}%
\begin{eqnarray}
\pi g_{2}\left( y\right) &=&1+\frac{1}{\beta _{1}}\int_{-1}^{1}dxg_{1}\left(
x\right) \left( b_{5}-b_{1}\right) \left( \frac{x}{\beta _{1}},\frac{y}{%
\beta _{2}}\right)  \label{g2} \\
&&+\frac{1}{\beta _{2}}\int_{-1}^{1}dy^{\prime }g_{2}\left( y^{\prime
}\right) \left( b_{6}+b_{4}-b_{2}\right) \left( \frac{y}{\beta _{2}},\frac{%
y^{\prime }}{\beta _{2}}\right) ,  \notag
\end{eqnarray}%
with%
\begin{equation}
\left( b_{n}-b_{m}\right) \left( x_{1},x_{2}\right) =b_{n}\left(
x_{1},x_{2}\right) -b_{m}\left( x_{1},x_{2}\right) ,
\end{equation}%
and%
\begin{equation}
b_{n}\left( x_{1},x_{2}\right) =\frac{n/4}{\left( n/4\right) ^{2}+\left(
x_{1}-x_{2}\right) ^{2}}.
\end{equation}%
The normalization conditions (\ref{n1}) and (\ref{n2}) are now
\begin{eqnarray}
\beta _{1} &=&\frac{2\gamma }{p}\int_{-1}^{1}g_{1}\left( x\right) dx,
\label{beta1} \\
\beta _{2} &=&\frac{4\gamma }{1-p}\int_{-1}^{1}g_{2}\left( y\right) dy.
\label{beta2}
\end{eqnarray}%
From (\ref{i energy}) the ground state energy per atom for the homogeneous
1D spinor gas is
\begin{equation}
\varepsilon ^{\hom }\left( n,\gamma ,p\right) =\frac{E}{N}=\frac{\hbar
^{2}n^{2}}{2m}e\left( \gamma ,p\right) ,  \label{varepsilon}
\end{equation}%
with
\begin{equation}
e\left( \gamma ,p\right) =e_{1}\left( \gamma ,p\right) +e_{2}\left( \gamma
,p\right) +e_{b}\left( \gamma ,p\right)  \label{e function}
\end{equation}%
and%
\begin{eqnarray}
e_{1}\left( \gamma ,p\right) &=&\frac{8\gamma ^{3}}{\beta _{1}^{3}}%
\int_{-1}^{1}x^{2}g_{1}\left( x\right) dx,  \label{e1} \\
e_{2}\left( \gamma ,p\right) &=&\frac{16\gamma ^{3}}{\beta _{2}^{3}}%
\int_{-1}^{1}y^{2}g_{2}\left( y\right) dy,  \label{e2} \\
e_{b}\left( \gamma ,p\right) &=&-\frac{\gamma ^{2}\left( 1-p\right) }{4}.
\label{eb}
\end{eqnarray}%
Here (\ref{e1}) and (\ref{e2}) can be solved numerically with the
combination of integral equations (\ref{g1}), (\ref{g2}) and the
normalization (\ref{beta1}), (\ref{beta2}). The chemical potentials are
taken as the derivatives of (\ref{varepsilon}) as%
\begin{eqnarray}
\mu _{1}^{\hom }\left( n,\gamma ,p\right) &=&\frac{\partial (n\varepsilon
^{\hom })}{\partial n_{1}}=\frac{\hbar ^{2}n^{2}}{2m}f_{1}\left( \gamma
,p\right) ,  \label{mu1} \\
\mu _{2}^{\hom }\left( n,\gamma ,p\right) &=&\frac{\partial (n\varepsilon
^{\hom })}{\partial n_{2}}=\frac{\hbar ^{2}n^{2}}{2m}f_{2}\left( \gamma
,p\right) ,  \label{mu2}
\end{eqnarray}%
where%
\begin{eqnarray}
f_{1}\left( \gamma ,p\right) &=&3e-\gamma \frac{\partial e}{\partial \gamma }%
+\left( 1-p\right) \frac{\partial e}{\partial p},  \label{f1} \\
f_{2}\left( \gamma ,p\right) &=&2\left( 3e-\gamma \frac{\partial e}{\partial
\gamma }-p\frac{\partial e}{\partial p}\right) .  \label{f2}
\end{eqnarray}

We see that, when $p=1$, the system is in a pure ferromagnetic phase with
spin-polarized bosons. Here $e\left( \gamma ,p=1\right) $ coincides with $%
e\left( \gamma \right) $ in the L-L model of scalar bosons \cite{Lieb1963}
with interaction parameter $2\gamma $. When $p=0$, all the bosons form pairs
and the system is in a pure antiferromagnetic phase. In the limiting case of
$\gamma =0$, the system reduces to free bosons with $e\left( \gamma
=0,p\right) =0$. When $\gamma $ is very strong, the integral equations (\ref%
{g1}) and (\ref{g2}) give $g_{1}\left( x\right) \approx 1/2\pi $ and $%
g_{2}\left( x\right) \approx 1/\pi $, then
\begin{equation}
e\left( \gamma \rightarrow +\infty ,p\right) \approx \frac{\pi ^{2}p^{3}}{3}+%
\frac{\pi ^{2}\left( 1-p\right) ^{3}}{48}-\frac{\gamma ^{2}\left( 1-p\right)
}{4}.
\end{equation}%
The energy per unit length in the strong interaction case%
\begin{equation}
\frac{E}{L}\approx \frac{\hbar ^{2}}{2m}\left( \frac{\pi ^{2}n_{1}^{3}}{3}+%
\frac{\pi ^{2}n_{2}^{3}}{6}-\frac{c^{2}n_{2}}{8}\right),
\end{equation}%
is composed of three parts: the energy density of $N_{1}$ free fermions with
mass $m$ in Ref. \cite{Hongmei2012}, the energy density of $N_{2}$ free
composite fermions with mass $2m$, and the binding energy $N_{2}\epsilon
_{b}/L$ of the composite fermions. This shows that stronger interactions
favors the forming of boson-boson pairs. The chemical potentials in this
case are%
\begin{eqnarray}
\mu _{1}^{\hom }\left( \gamma \rightarrow +\infty ,p\right) &\approx &\frac{%
\hbar ^{2}}{2m}\pi ^{2}n_{1}^{2},  \label{mu1 at strong} \\
\mu _{2}^{\hom }\left( \gamma \rightarrow +\infty ,p\right) &\approx &\frac{%
\hbar ^{2}}{2m}\left( \frac{\pi ^{2}}{2}n_{2}^{2}-\frac{c^{2}}{8}\right) .
\label{mu2 at strong}
\end{eqnarray}

\begin{figure}[tbp]
\includegraphics[width=0.45\textwidth]{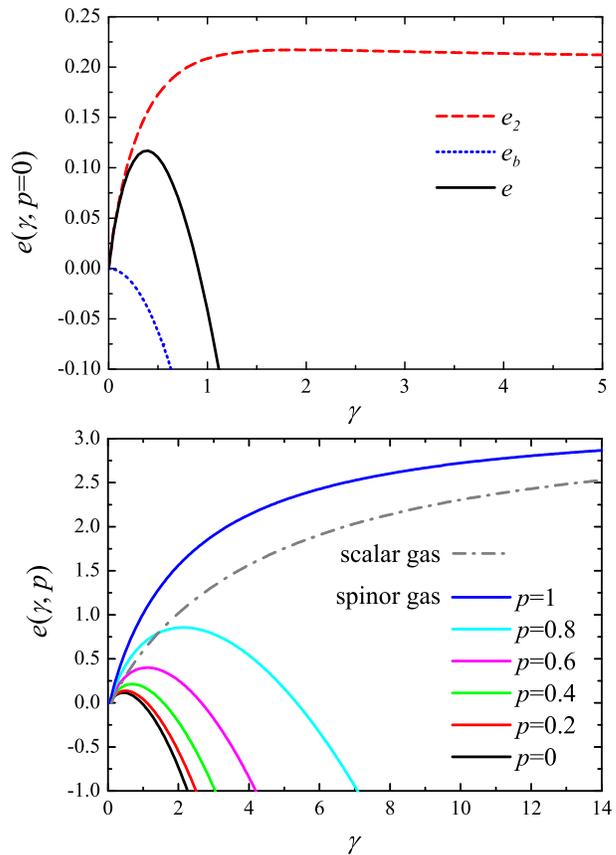}
\caption{ (Color Online) The function $e\left( \protect\gamma ,p\right) $
for the ground state energy of 1D homogeneous spin-1 bosons. In the upper
panel, $p=0$. In the lower panel the real lines are for $%
p=0,0.2,0.4,0.6,0.8,1$ from bottom to top and the dash-dotted line is for
scalar bosons.}
\label{e}
\end{figure}

In Fig. \ref{e}, we plot the interaction dependence of energy density $%
e\left( \gamma ,p\right) $ for various polarization $p$ and compare it with
the case of scalar bosons. The upper figure is for $p=0$, i.e., all the
bosons form the pairs. Clearly we have $e_{1}=0$ and $e_{2}\left( \gamma
\right) $ (red dashed line), which is obtained numerically from the
combination of (\ref{g2}), (\ref{beta2}) and (\ref{e2}), increases linearly
along with $\gamma $\ for $\gamma \ll 1$ and approaches slowly to a constant
value $\pi ^{2}/48$\ for $\gamma \gg 1$. $e_{2}\left( \gamma \right) $ and
the inverse parabola function $e_{b}\left( \gamma \right) =-\gamma ^{2}/4$
together dictate that $e\left( \gamma \right) $ is not monotonic and has a
maximum value $0.1169$\ at $\gamma =0.4$. The lower figure is for $e\left(
\gamma ,p\right) $ with $p=0,0.2,0.4,0.6,0.8,1$ corresponding to the real
lines from bottom to top. It shows that the energy function $e\left( \gamma
,p\right) $ increases along with $p$ for the same parameter $\gamma $. For $%
p=1$, i.e., pure ferromagnetic phase case, $e\left( \gamma \right) $
increases monotonously to constant $\pi ^{2}/3$ which is exactly the
asymptotic value of scalar bosons (gray dash-dotted line).

\subsection{ Kohn-Sham equations for trapped system}

Now we consider the spin-1 bosons in a harmonic trap $V_{ext}\left( x\right)
=m\omega ^{2}x^{2}/2$ by means of the DFT
theory based on the theorems of Hohenberg, Kohn and Sham. The theory enables
us to deal with the energy and the density profile of inhomogeneous system
in the ground state. According to the Hohenberg-Kohn theorem I of DFT, the
ground-state density of a bound system of interacting particles in some
external potential determines this potential uniquely. Denote now the space
dependent densities of particle I and particle II as $n_{1}\left( x\right) $
and $2n_{2}\left( x\right) $, respectively. The total density is then $%
n\left( x\right) =n_{1}\left( x\right) +2n_{2}\left( x\right) $ and the
number of the particles are conserved separately according to
\begin{eqnarray}
\int n_{1}\left( x\right) dx &=&N_{1},  \label{n1N1} \\
\int n_{2}\left( x\right) dx &=&N_{2}.  \label{n2N2}
\end{eqnarray}%
The ground-state energy is a functional of the densities $E_{0}\left[
n_{1}\left( x\right) ,n_{2}\left( x\right) \right] $. It can be decomposed
as kinetic energy functional of a reference noninteracting system $T^{ref}%
\left[ n_{1}\left( x\right) ,n_{2}\left( x\right) \right] $, external
potential energy functional $E_{ext}\left[
n_{1}\left( x\right) ,n_{2}\left( x\right) \right]$
and KS energy functional $E_{KS}\left[ n_{1}\left(
x\right) ,n_{2}\left( x\right) \right] $ involving the interactions, i.e.,
\begin{eqnarray}
E_{0}\left[ n_{1},n_{2}\right] &=&T^{ref}\left[ n_{1},n_{2}\right] +
E_{ext}\left[n_{1},n_{2} \right]  \notag \\
&&+E_{KS}\left[ n_{1},n_{2}\right] .  \label{energy functional 1}
\end{eqnarray}

We introduce two orthogonal and normalized Bose orbital functionals $\phi
_{1}\left( x\right) $ and $\phi _{2}\left( x\right) $ to express the
densities as
\begin{eqnarray}
n_{1}\left( x\right) &=&N_{1}\phi _{1}^{\ast }\left( x\right) \phi
_{1}\left( x\right) ,  \label{density1} \\
n_{2}\left( x\right) &=&N_{2}\phi _{2}^{\ast }\left( x\right) \phi
_{2}\left( x\right) .  \label{density2}
\end{eqnarray}%
The kinetic energy functional is written as
\begin{eqnarray}
T^{ref}\left[ n_{1},n_{2}\right] &=&-N_{1}\int dx\phi _{1}^{\ast }\left(
x\right) \frac{\hbar ^{2}}{2m}\frac{d^{2}}{dx^{2}}\phi _{1}\left( x\right)
\notag \\
&&-2N_{2}\int dx\phi _{2}^{\ast }\left( x\right) \frac{\hbar ^{2}}{2m}\frac{%
d^{2}}{dx^{2}}\phi _{2}\left( x\right)   \label{Tref}
\end{eqnarray}%
and the external potential energy functional is simply
\begin{equation}
E_{ext}\left[
n_{1},n_{2} \right]=\int
dxV_{ext}\left( x\right) n\left( x\right) .
\end{equation}
Note for each part of bosons we introduce a single orbital functional
assuming that the bosons are in a quasi-condensate state. This is different
from the system of fermions for which the number of orbital functional is
decided by the number of fermions. In Ref. \cite{Hongmei2012} we have
indicated the validity of single Bose orbital functional in DFT. It gives
the density profile of 1D bosons varying from a standard Gaussian shape for
weak interaction to a half-ellipse profile for strong interaction. The
density profile for strong interaction is consistent with that of
noninteracting fermions except the density oscillations. In the limit of
large particle number, the difference between the oscillating and
non-oscillating profiles becomes imperceptible.

$E_{KS}\left[ n_{1},n_{2}\right] $ includes all the contribution of the
interaction energies. Sometimes it is partitioned as Hartree-Fock energy
(i.e., the mean field approximation of the interaction energy) and exchange
correlation energy \cite{HohenbergKohn,Xianlong,Hongmei2012}. Following the
way in \cite{Nunes,Kim,Brand,Yajiang2009}, we here treated it as an entity
with the LDA, that is, the system can be assumed locally equilibrium at each
point $x$ in the external trap, with local energy per atom provided by the
homogenous interactional system. Thus the interaction energy functional $%
E_{KS}\left[ n_{1},n_{2}\right] $ can be formulated as%
\begin{equation}
E_{KS}\left[ n_{1},n_{2}\right] \approx \int dxn\left( x\right) \varepsilon
^{\hom }\left[ n_{1}\left( x\right) ,n_{2}\left( x\right) \right]
\label{EKS}
\end{equation}%
where the densities $n_{1}\left( x\right) ,n_{2}\left( x\right) $ are taken
at point $x$ and $\varepsilon ^{\hom }\left[ n_{1}\left( x\right)
,n_{2}\left( x\right) \right] $ takes the form of Eq. (\ref{varepsilon}).
Note that both the L-L parameter $\gamma \left( x\right) =mg/\hbar ^{2}n\left(
x\right) $ and the polarization parameter $p\left( x\right) =n_{1}\left(
x\right) /n\left( x\right) $ are space dependent now. With the explicit form
of three terms in (\ref{energy functional 1}), the ground state energy
functional are
\begin{eqnarray}
&&E_{0}\left[ n_{1}\left( x\right) ,n_{2}\left( x\right) \right]  \notag \\
&=&N_{1}\int dx\phi _{1}^{\ast }\left( x\right) \left[ -\frac{\hbar ^{2}}{2m}%
\frac{d^{2}}{dx^{2}}+\frac{1}{2}m\omega ^{2}x^{2}\right] \phi _{1}\left(
x\right)  \notag \\
&&+2N_{2}\int dx\phi _{2}^{\ast }\left( x\right) \left[ -\frac{\hbar ^{2}}{2m%
}\frac{d^{2}}{dx^{2}}+\frac{1}{2}m\omega ^{2}x^{2}\right] \phi _{2}\left(
x\right)  \notag \\
&&+\int dxn\left( x\right) \varepsilon ^{\hom }\left[ n_{1}\left( x\right)
,n_{2}\left( x\right) \right] .  \label{energy functional 2}
\end{eqnarray}

Hohenberg-Kohn theorem II guarantees that the ground-state density
distributions are determined by variationally minimizing $E_{0}$ with
respect to $n_{1}\left( x\right) $ and $n_{2}\left( x\right) $ \cite%
{HohenbergKohn}. That is equivalent to minimize the free-energy functional $%
F=E_{0}-N_{1} \epsilon _{1}-2N_{2}\epsilon _{2}$ with respect to $\phi
_{1}^{\ast }$ and $\phi _{2}^{\ast }$, where the Lagrange multipliers $%
\epsilon _{1},\epsilon _{2}$ are introduced to conserve $N_{1}$ and $2N_{2}$%
. Then we can get the KS equations%
\begin{eqnarray}
&&\left( -\frac{\hbar ^{2}}{2m}\frac{d^{2}}{dx^{2}}+\frac{1}{2}m\omega
^{2}x^{2}+\mu _{1}^{\hom }\left[ n_{1}\left( x\right) ,n_{2}\left( x\right) %
\right] \right) \phi _{1}\left( x\right)  \notag \\
&=&\epsilon _{1}\phi _{1}\left( x\right) ,  \label{KS1}
\end{eqnarray}%
\begin{eqnarray}
&&\left( -\frac{\hbar ^{2}}{2m}\frac{d^{2}}{dx^{2}}+\frac{1}{2}m\omega
^{2}x^{2}+\frac{1}{2}\mu _{2}^{\hom }\left[ n_{1}\left( x\right)
,n_{2}\left( x\right) \right] \right) \phi _{2}\left( x\right)  \notag \\
&=&\epsilon _{2}\phi _{2}\left( x\right) ,  \label{KS2}
\end{eqnarray}%
where the chemical potentials $\mu _{1}^{\hom },$ $\mu _{2}^{\hom }$ of the
homogeneous gas for densities $n_{1},n_{2}$ at $x$ are given by Eqs. (\ref%
{mu1}) and (\ref{mu2}). With the eigenvalues of (\ref{KS1}) and (\ref{KS2}),
the ground state energy can be expressed as
\begin{eqnarray}
E_{0} &=&N_{1}\epsilon _{1}+2N_{2}\epsilon _{2}+\int dxn\left( x\right)
\varepsilon ^{\hom} \left( x\right)  \notag \\
&&-\int n_{1}\left( x\right) \mu _{1}\left( x\right) dx-\int n_{2}\left(
x\right) \mu _{2}\left( x\right) dx.  \label{Ground state energy}
\end{eqnarray}%
With the exactly solved $\varepsilon ^{\hom }$ for different $n_{1},n_{2}$
at $x$, we can solve the KS equations (\ref{KS1}) and (\ref{KS2}) together
with the definition of orbital functional (\ref{density1}) and (\ref%
{density2}) to find the density distributions $n_{1}\left( x\right) $ and $%
n_{2}\left( x\right) $ and then calculate the ground-state energy $E_{0}$
from Eq. (\ref{Ground state energy}).

\begin{figure}[tbp]
\includegraphics[width=0.45\textwidth]{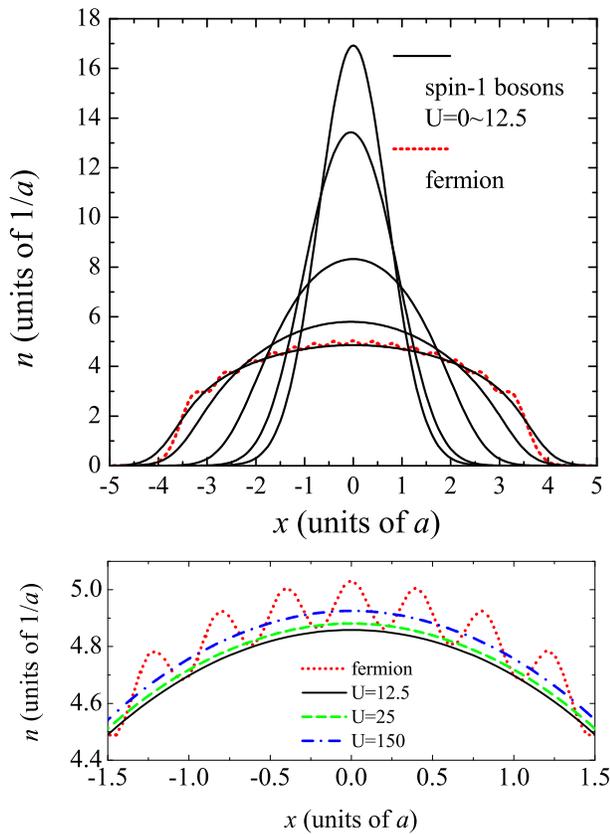}
\caption{(Color Online) Density distribution of 1D trapped spin-1 bosons
with $N=30$ and $P=0$. In the upper figure, five black solid lines from top
to bottom are respectively for spinor bosons with $U=0,0.1,0.5,2,12.5$. Red
dotted line is for non-interacting Fermions with mass $2m$. The lower figure 
shows the details of the peak density for even stronger interactions 
$U=12.5,25,150$. We observe a slightly rising of the peak after $U>12.5$.}
\label{densityp0}
\end{figure}

\begin{figure}[tbp]
\includegraphics[width=0.45\textwidth]{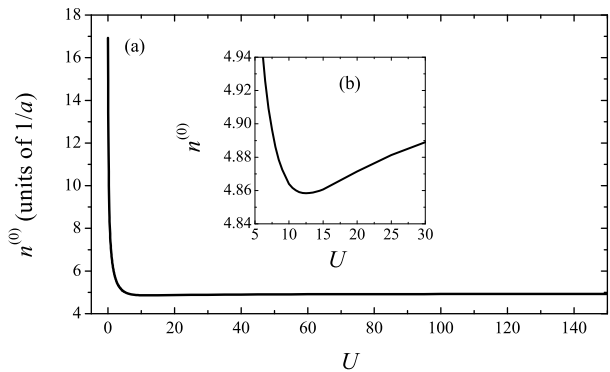}
\caption{Non-monotonicity of the central density $n^{(0)}$ of 1D trapped
spin-1 bosons with increasing $U$ for $N=30$ and $P=0$. The exists a minimum value 
of $n^{(0)}$ at $U=12.5$ which is shown in the inset.}
\label{densityP0x0}
\end{figure}

We now discuss the KS equations for the limiting cases of weak and strong interaction. When
there is no interaction, KS equations correctly reduce to the 1D Schr\"{o}%
dinger equation of simple harmonic oscillator. The Bose density profiles take
the standard Gaussian shape%
\begin{equation}
n_{1,2}\left( x\right) =\frac{N_{1,2}}{a\sqrt{\pi }}\exp \left(
-x^{2}/a^{2}\right) .  \label{n1,2 weak}
\end{equation}%
When the interaction is strong, the kinetic energies in (\ref{KS1}) and (\ref%
{KS2}) can be ignored, we have the TFA equations
\begin{eqnarray}
\frac{1}{2}m\omega ^{2}x^{2}+\mu _{1}\left( x\right) &=&\mu _{1}^{0},
\label{TFA1} \\
\frac{1}{2}m\omega ^{2}x^{2}+\frac{\mu _{2}\left( x\right) }{2} &=&\frac{\mu
_{2}^{0}}{2}.  \label{TFA2}
\end{eqnarray}%
The values of $\mu _{1}^{0}$ and $\mu _{2}^{0}$ are fixed by the
normalization conditions (\ref{n1N1}) and (\ref{n2N2}). Based on (\ref{TFA1}%
) and (\ref{TFA2}) Kuhn \textit{et al.} give the ground state phase diagram
and find that the singlet pairs and unpaired bosons may form a two-component
Luttinger liquid in the strong coupling regime \cite{Kuhn2012}. When the
interaction approaches infinitely strong, with the limit values of chemical potential (%
\ref{mu1 at strong}, \ref{mu2 at strong}) and the normalization conditions (%
\ref{n1N1}, \ref{n2N2}), we can solve (\ref{TFA1}, \ref{TFA2}) to obtain the following
half-ellipse-like density profiles
\begin{equation}
n_{1,2}\left( x\right) \approx \frac{\sqrt{2N_{1,2}-\left( x/a_{1,2}\right)
^{2}}}{\pi a_{1,2}},  \label{TFA n2}
\end{equation}%
where $a_{1}=a$ and $a_{2}=\sqrt{\hbar /2m\omega }$. We see that the density
of particle I is just that of $N_{1}$ noninteracting harmonically trapped
fermions with mass $m$. The density of particles II is that of $N_{2}$
noninteracting fermions with mass $2m$. It shows that for infinitely strong
interaction, the property of particles II approaches the TG gas of Bose-Bose
pairs. Resorting to the Bose-Fermi mapping method \cite%
{Deuretzbacher2008,Girardeau2011,Girardeau1960}, we may map the densities of
paired and unpaired components exactly to those of non-interacting Fermions%
\begin{equation}
n_{1,2}\left( x\right) =\frac{1}{a_{1,2}\sqrt{\pi }}\exp \left(
-x^{2}/a_{1,2}^{2}\right) \sum_{{l=0}}^{N_{1,2}-1}\frac{H_{l}^{2}\left(
x/a_{1,2}\right) }{2^{l}l!},  \label{TG}
\end{equation}%
where $H_{l}\left( x\right) $ is the Hermite polynomials.

\section{Numerical results}

We introduce the interacting parameter $U=g/a\hbar \omega $. The
space-dependent Lieb-Liniger parameter is expressed as $\gamma \left(
x\right) =U/an\left( x\right) $. For the 1D spin-1 bosons with given $N$, $P$
and $U$, we present the numerical results for densities $n_{1}$ and $n_{2}$ of particle I and II
obtained by solving the KS equations (\ref{KS1}) and ({\ref{KS2}) with
the iteration method. The ground state energy $E_{0}$ can be obtained via
the relation (\ref{Ground state energy}). The numerical results are summarized in Figs. \ref{densityp0}%
-\ref{energyPa}.

\subsection{$P=0$ system}

We first study the system with $N=30$ and $P=0$. 
In this fully paired case, the length, density and energy
are in units of $a$, $1/a$ and $\hbar \omega $ respectively.
Fig. \ref{densityp0}
provides an understanding of how the density profiles change along
with the interaction parameter for $U=0,0.1,0.5,2,12.5$ in the upper
figure and $U=12.5,25,150$ in the lower figure. With the increasing of
$U$, the total density profile varies from a standard Gaussian-like shape characterizing
the distribution of non-interacting Bose gas to a half-ellipse shape
indicating the distribution of non-interacting Fermi gas. Interestingly we find that
the density profile of $N$ spin-1 bosons with mass $m$ for strongly
interacting case e.g. $U=150$ overlaps that of $N/2$ noninteracting
fermions with mass $2m$, except the emergence of the density oscillation
in the Fermionic case. It means that for strong interaction
all atoms in the anti-ferromagnetic spin-1 bosons are paired with each other,
behaving like the TG gas of Boson pairs.

We surprisingly notice that the density profile does not change with the interaction
parameter monotonically. The full tendency of the central density $n^{(0)}$ (density
at the trap center $x=0$) can be seen in Fig. \ref{densityP0x0}
and the inset shows in detail the non-monotonicity of $n^{(0)}$.
We find that central density firstly decreases to a minimum value $n^{(0)}_{\min }=4.858/a$
at $U=12.5$ before reaching the constant central density of non-interacting
fermions for large $U$. The result can be understood as the competition
between the repulsive density-density interaction and the anti-ferromagnetic
spin-exchange interaction with equal strength.
The repulsive density-density interaction tends to increase the distance
between bosons while the anti-ferromagnetic spin-exchange
interaction leads essentially attraction between $s=\pm 1$ or $s=0$ bosons. For
$U<12.5$, the prevailing repulsions among bosons tend to broaden them to
wider space area and reduce the density of bosons in the trap center
to a minimum. The anti-ferromagnetic effect is prominent
for $U>12.5$ which contracts the bosons slightly.

\begin{figure}[tbp]
\includegraphics[width=0.45\textwidth]{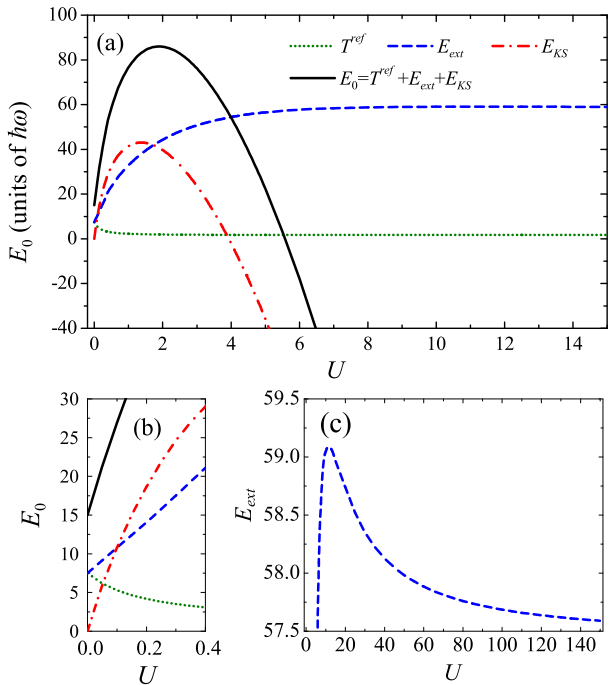}
\caption{(Color Online) (a) Evolution of the ground state energy
$E_{0}$ and all contributed energy terms of 1D trapped spin-1 bosons
with increasing $U$ for $N=30$ and $P=0$. (b) the details of energies in the weak
interaction regime. (c) the details of external potential energy.}
\label{COMenergyp0}
\end{figure}

Fig. \ref{COMenergyp0} describes the evolution of the ground state energy
$E_{0}$ and all contributed energy terms in Eq. (\ref{energy functional 1}) as
a function of $U$. We can see the whole trend in (a) and
the details in the mean field regime in (b). It shows
that the kinetic energy $T^{ref}$ decreases slowly to a constant
energy as the result of interactions restraining the movement of atoms.
However, the external potential energy $E_{ext}$ increases due to
the atoms occupy wider regime of the trap. In correspondence to the
non-monotonicity of $n^{(0)}$, $E_{ext}$ shows non-monotonicity too,
which can be seen clearly in Fig. \ref{COMenergyp0}(c). Close to
the critical interaction value of $U=12.5$ for $n^{(0)}_{min}$ in Fig.
(\ref{densityP0x0}c), $E_{ext}$ reaches its maximum $E_{ext}^{\max }=59.092\hbar \omega
$ at $U=11.5$. But the non-monotonicity of $n^{(0)}$ doesn't develop visible effects on
other energy terms. Analogous to the ground state energy functional $e\left(
\gamma ,p=0\right) $ for homogeneous spin-1 bosons (see the upper panel of
Fig. \ref{e}), the KS energy $E_{KS}$, i.e. the interaction energy, increases linearly in weak
interaction regime and approaches its peak value
$E_{KS}^{\max }=43.033\hbar \omega $ at $U=1.4$, corresponding to an axial
L-L interaction parameter $\gamma ^{(0)}=0.222$ ($\gamma $ at the trap center),
followed by a monotonously decreasing in strong interaction regime.
The contributions from $T^{ref}$, $E_{ext}$ and $E_{KS}$
together establish the ground state energy $E_{0}$ to increase to a maximum $%
E_{0}^{\max }=85.994\hbar \omega $ at $U=1.9$, corresponding to
$\gamma ^{(0)}=0.324$, and then decrease monotonously. We notice that
the value $\gamma ^{(0)}=0.324$ for $E_{0}^{\max }$ are close to $%
\gamma =0.4$ for the maximal $e$ in the upper panel of Fig. \ref{e}. We
understand that the competition between the repulsion among paired bosons
and the binding energy of the pair gives the peak of $E_{0}$ at $U=1.9$ for the
same reason as in the homogeneous case, while that between the repulsive
density-density interaction and the anti-ferromagnetic
spin-exchange interaction gives the non-monotonicity of
$n^{(0)}$ at $U=12.5$.

\subsection{$P\neq 0$ system}

For $P=1$, none of the bosons can form pair and the system reduces to the
scalar Bose gas with density-density interaction $2g$, whose ground state
properties have been studied by means of DFT in Ref. \cite{Hongmei2012}.
With the increasing of $U$, the density distribution and energy of the
system approach those of a single-component TG gas.

\begin{figure*}[t]
\includegraphics[width=0.9\textwidth]{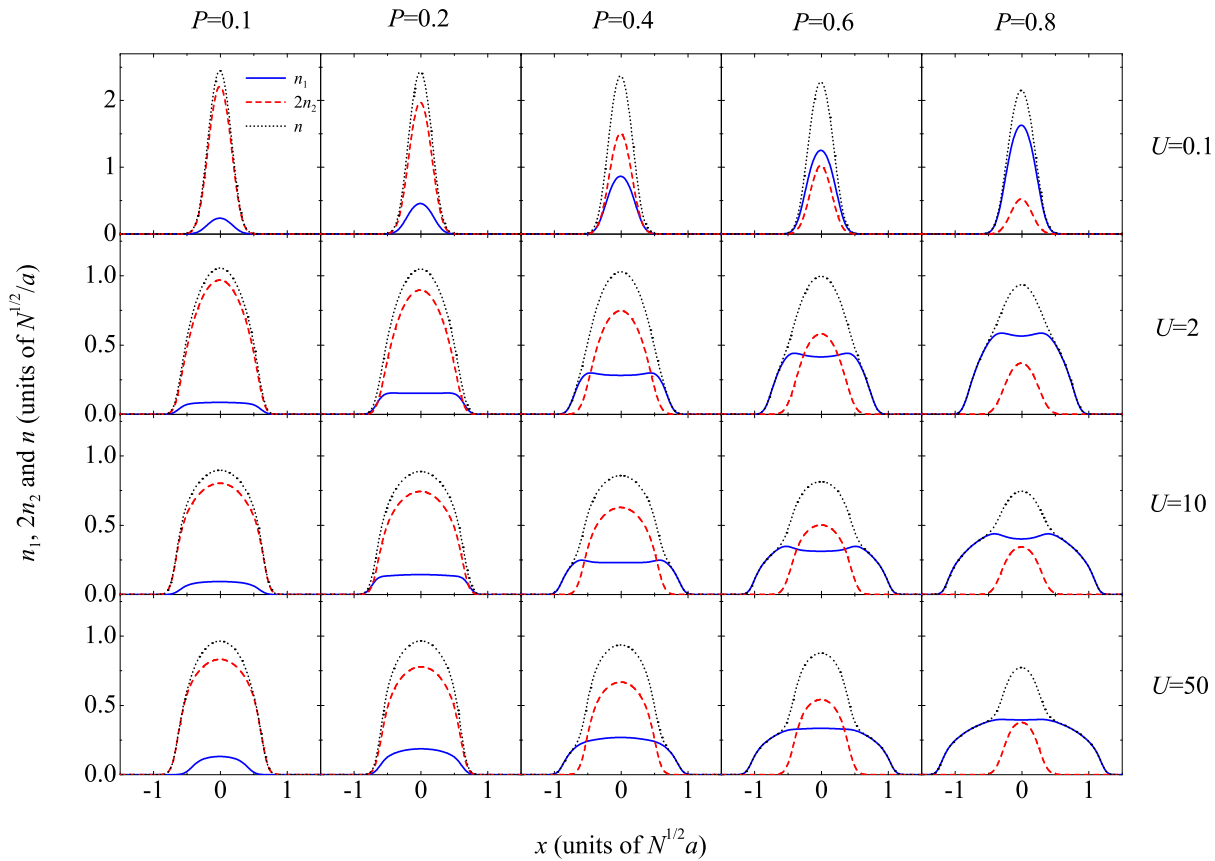}
\caption{(Color Online) Density distributions of 1D trapped spin-1
bosons at $P=0.1,0.2,0.4,0.6,0.8$ (panels from left to right) and $U=0.1,2,10,50$
(panels from top to bottom). Blue solid lines denote the densities for polarized particles $n_1$,
red dashed lines denote the densities for paired bosons $2n_2$, and black dotted lines denote
the total densities $n$. }
\label{densityPaUa}
\end{figure*}

Here we study the interesting situation of a partially polarized
spinor gas with total polarization $0<P<1$. 
In this case the length, density and energy
are in units of $N^{1/2}a$, $N^{1/2}/a$ and $\hbar \omega $ respectively.
In Fig. \ref{densityPaUa} we illustrate the density
of polarized particles $n_{1}$ (blue solid lines), density of paired bosons
$2n_{2}$ (red dashed lines) and the total density $n$ (black dotted lines)
for various polarization $P=0.1,0.2,0.4,0.6,0.8$ (the figures in the columns from left to
right) and interaction $U=0.1,2,10,50$ (the figures in the rows from up to down).
Horizontal view shows that the density profile $n_{1}$ expands gradually with the
increasing of $P$ accompanied by the corresponding shrinking of $2n_{2}$. These two
densities together result in a slightly decreasing of the peak density $n^{(0)}$
for increasing $P$. An analysis of vertical scope for increasing interaction
parameter $U$ (note that we use different vertical axis scale in the weak interaction
case $U=0.1$) tells us that the density $2n_{2}$ changes smoothly from the
Gaussian distribution of non-interacting Bose gas to the half-ellipse
distribution of TG gas of Bose-Bose pairs. The density peak
is located in the center of the trap for all interaction strengths.
The behavior of density $n_{1}$ is, however, a little complicated.
Though $n_{1}$ shows Gaussian distribution for weak interaction ($U=0.1$),
bosons of Particle I tend to occupy wider space and start to be excluded
from the trap center for intermediate interaction ($U=2,10$). The single peak
profile of $n_{1}$ changes into the double-peak distribution. This reminds
us the partially phase separation of Bose-Fermi mixture with equal mass and
equal repulsive interaction \cite{Imambekov2006,Hongmei2012}.
In our case, the phase separation of paired and unpaired bosons
occurs in the system, i.e. the core area of the 1D spinor gas is filled with the
mixture of paired and unpaired bosons and in the outer region we find either
polarized bosons for $P>0.2$ or paired bosons for $P<0.2$. For even stronger 
interaction ($U=50$), the peak of $n_{1}$
returns back to the trap center and the density profile $n_{1}$ approaches
the half-ellipse distribution of TG gas. Nevertheless the evidence of phase separation
becomes more prominent in the strongly interaction case.

\begin{figure}[tbp]
\includegraphics[width=0.45\textwidth]{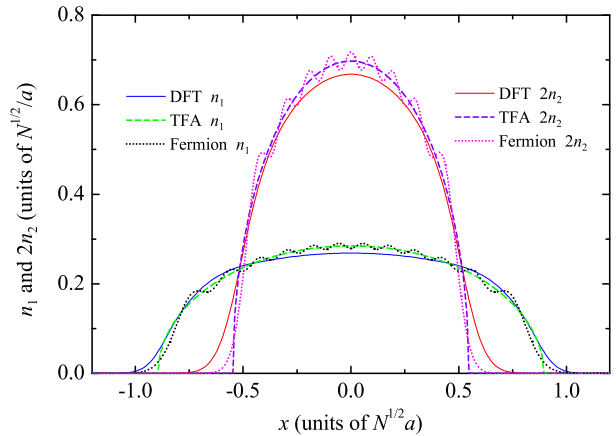}
\caption{(Color Online) Comparison of the density distribution of 1D trapped spin-1 bosons at 
$P=0.4$ in the case of strong interaction. The results for both $n_1$ and $2n_2$ are from DFT for 
$U=50$ (solid lines), TFA for $U=+\infty $ (dashed lines), and $N=30$ non-interacting fermions 
(dotted lines), respectively. Clear evidence is observed for the phase separation and in this case 
the surrounding wings are composed of polarized particles. The radius of vanishing density
from DFT is obviously larger than the TFA result.}
\label{compareP04U50}
\end{figure}

As a result the total density $n$ shows the overall Gaussian distribution
showing a fully mixing phase of $n_{1}$ and $2n_{2}$ at weak interaction
($U=0.1$). When the interaction increases, we find a bi-modal distribution
of the total density with particle II imposed on the top of particle I, i.e.
the mixed core of particles I and II is surrounded by two wings composed of
solely polarized bosons for large $P$ and large $U$ (see the several right-down
panels of Fig. \ref{densityPaUa}). Take the $P=0.4$ case as an example. We
compare the density plots for the two components in Fig. \ref{compareP04U50}.
The solid lines are our DFT results from the iteration solution of the KSE
Eqs. (\ref{KS1}) and ({\ref{KS2}), while the dashed lines are analytical TFA
results Eq. (\ref{TFA n2}) for infinitely strong interaction. Also shown are
the densities of non-interacting fermions Eq. (\ref{TG}) according to the
Bose-Fermi mapping (dotted lines) for $N=30$. We identify readily the density oscillations
with 12 peaks in the polarized component $n_1$ in the TG limit and 9 peaks in
the paired component $n_2$ representing the TG gas of 9 pairs of bosons.

\begin{figure}[tbp]
\includegraphics[width=0.45\textwidth]{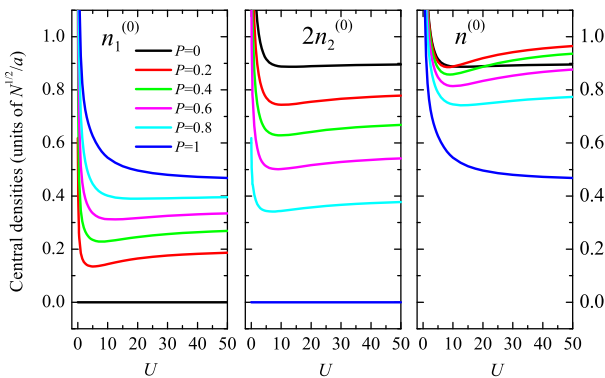}
\caption{(Color Online) Evolution of central densities $n_{1}^{(0)}$, $%
n_{2}^{(0)}$ and $n^{(0)}$ (panels from left to right) of trapped spin-1
bosons with increasing interaction parameter $U$.  Non-monotonicity is
seen for all partially polarized cases $0<P<1$.}
\label{densityPax0}
\end{figure}

An interesting observation is the non-monotonicity of the central density value
$n^{(0)}$ which already occurs in the fully paired case $P=0$. We take a close look
at the rise-and-fall of the peak values for polarized bosons, paired bosons, and the total
density. Fig. \ref{densityPax0} shows the evolution of central densities with increasing
interaction parameter $U$ for different $P$. For the central density of
polarized particles $n_{1}^{(0)}$, we find it is a constant zero at $P=0$ since there
are no Particle I in the gas and a monotonically decreasing curve at $P=1$ corresponding
to the scalar Bose gas. In all partially polarized case $0<P<1$, the peak values
rapidly decrease to a minimum $n_{1\min }^{(0)}$ and then gradually
approach to a constant for strong interaction. The minimum is seen to move toward
the stronger interaction direction with increasing $P$ and finally disappear for $P=1$.
On the contrary, the central density of paired particles $n_{2}^{(0)}$ is a constant zero at
$P=1$ since there are no paired bosons in the gas and a seemingly monotonically
decreasing curve at $P=0$ corresponding to the fully paired bosons.
Yet we know from the result of $P=0$ in Fig. \ref{densityP0x0} there indeed exists
a minimum due to the competition between the density interaction and spin exchange.
The competition persists here for all partially polarized cases,
leading to similar shallow low-lying areas, whose minimum is seen to move toward
the weaker interaction direction with increasing $P$ as shown in the middle panel
of Fig. \ref{densityPax0}. The total central density comes from the combination
of these two terms $n^{(0)}=n_{1}^{(0)}+2n_{2}^{(0)}$. The right panel shows that the
minimum hollow for the total central density remains for all partially or fully polarized cases
$0\le P<1$ and the tail of $n^{(0)}$ in strong interaction limit firstly goes up slightly
followed by a nearly linear decreasing  for increasing $P$. This non-monotonicity
of the total central density in strong interaction case ($U=50$) is depicted in detail in
Fig. \ref{compareDENSITYx0} for increasing $P$, together with the monotonic behavior of
the central density for polarized bosons (down-ward) and that for paired bosons (up-ward).
The results from DFT are compared with the analytical curves from the TFA.
We find that in the case of $U=50$ the central densities for all polarized situation are already
very close to the limiting value in TFA, which are obtained from eq. (\ref{TFA n2}) as
\begin{eqnarray}
n_1^{(0)}&=&\frac{\sqrt{2NP}}{\pi a},  \notag \\
2n_2^{(0)}&=&\frac{2\sqrt{2N(1-P)}}{\pi a}, \notag \\
n^{(0)}&=&\frac{\sqrt{2N}(\sqrt{P}+2\sqrt{1-P})}{\pi a}.
\end{eqnarray}

\begin{figure}[tbp]
\includegraphics[width=0.45\textwidth]{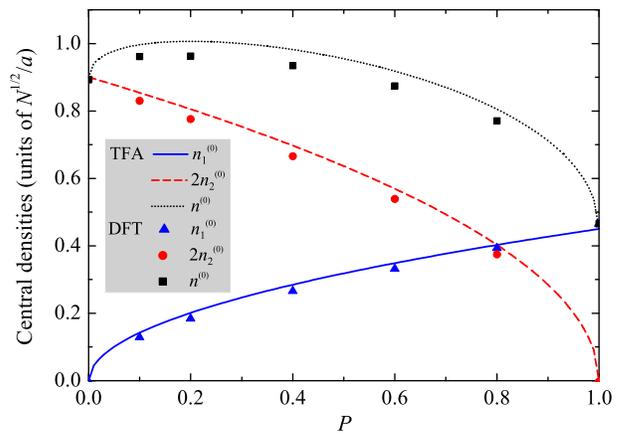}
\caption{(Color Online) Central densities $n_{1}^{(0)}$, $%
n_{2}^{(0)}$ and $n^{(0)}$ as a function of $P$ for 1D
trapped spin-1 bosons in the strong interaction case. 
The results of DFT for $U=50$ denoted by lines are in agreement with those in TFA 
for $U=+\infty $ denoted by symbols. }
\label{compareDENSITYx0}
\end{figure}

We may go further into the detailed quantum phases of the spinor gas
by defining the radii of the vanishing densities for the two components.
Fig. \ref{compareRADIUS} shows the phase diagram as a function of the
global polarization $P$ for $U=50$, where the axial radii of the ensemble of
the polarized and paired components are extracted from the numerical
result of the density profiles from the KSE. Without loss of generality, we set
the threshold values of the vanishing scaled density as $0.02$ in the numerical
simulation. The intersection of these two radii gives the boundaries which
divided the phase plane into three quantum phases: spin-singlet-paired
bosons S, ferromagnetic spin-aligned bosons F, mixed phase of the pairs
and unpaired bosons M, while V stands for the vacuum. At low polarization
a partially polarized region forms at the trap center, the radius of which
increases with increasing polarization. At a critical polarization $P_c$, the
partially polarized region extends to the edge of the cloud. When the
polarization increases further, the edge of the cloud becomes fully polarized.
This process can be evidently seen in the lowest panels in Fig. \ref{densityPaUa}.
Together shown in Fig. \ref{compareRADIUS} are the theoretical results in
TFA. According to eqs. (\ref{TFA n2}), the radii of the vanishing densities
are calculated as
\begin{eqnarray}
R_1&=&\sqrt{2NP}a, \notag \\
R_2&=&\sqrt{\frac{N(1-P)}{2}}a
\end{eqnarray}
When the two radii equal to each other, we find the critical polarization is
$P_c=0.2$ and the critical radius is $R_c=0.63$. The DFT results for the
radii are apparently larger than the TFA
estimation for both polarized and fully paired bosons, which can be understood
easily from the extension of the tail of the density profile into outer region
(see Fig. \ref{compareP04U50}). Energetically this extension is due to the
kinetic term neglected in TFA. The critical polarization is in agreement with
the TFA crossing at slightly higher polarization $P_c\sim 0.23$ and larger radius
$R_c\sim 0.73$.

\begin{figure}[tbp]
\includegraphics[width=0.45\textwidth]{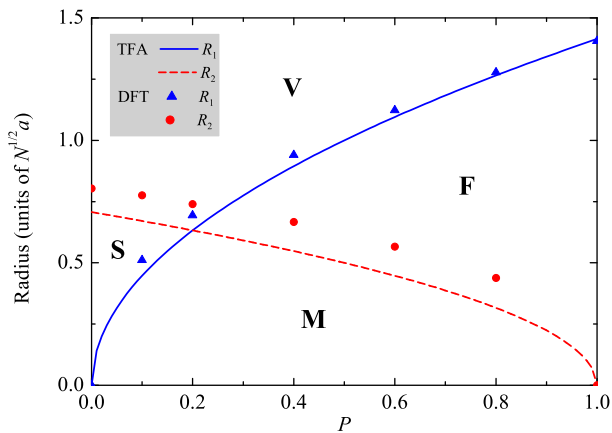}
\caption{(Color Online) R-P phase diagram of 1D trapped spin-1 bosons in terms of
the scaled density radii $R$ as the functions of total polarization $P$. Three
quantum phases are identified: spin-singlet-paired phase (S), ferromagnetic spin-aligned
phase (F) and mixed phase of the pairs and unpaired bosons (M). V stands for
vacuum. The results of DFT for $U=50$ ($n_{1}$, $2n_{2}<0.02$ are considered as
vanishing scaled densities) and analytical results in TFA for $U=+\infty $ are compared.}
\label{compareRADIUS}
\end{figure}

The phenomena of phase separation at strong interaction is consistent with
the results of Ref. \cite{Kuhn2012}. There they gave the ground
state phase diagram according to the evolvement of Thomas-Fermi radii
of $n_{1}$ and $n_{2}$ along with $P$ by the means of TBA
for strong interaction case. Our results, on the other hand, are valid for
systems in the whole interaction regime. Interestingly we found the
double-peak structure of the density of polarized bosons at intermediate interaction,
which is elusive from the method in \cite{Kuhn2012}. In a seminal experiment
on spin-imbalanced 1D two-component Fermi gas \cite{Liao2010},
phase separation of the two-spin mixture of ultracold $^6$Li atoms trapped
in an array of 1D tubes is reported. The partially polarized core is surrounded
by wings which are composed of either a paired or a polarized Fermi gas
depending on the polarization $P$. The pair mechanism in spinor gas is challenged
by the density repulsive interaction, which makes the phase separation more complicated.

\begin{figure}[!tbp]
\includegraphics[width=0.45\textwidth]{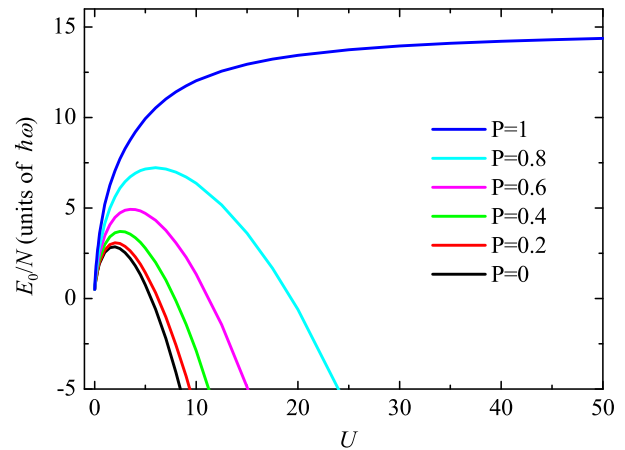}
\caption{(Color Online) The ground state energy per atom of 1D trapped
spin-1 bosons as functions of $U$ for different $P$. The lines are for
P=0,0.2,0.4,0.6,0.8,1 from bottom to top.}
\label{energyPa}
\end{figure}

Finally, just as $e(\gamma,p)$ shows the interaction dependence for various
polarizations in the homogeneous case and $E_0$ in Eq. (\ref{Ground state energy})
gives the change with interaction in the fully paired case, we illustrate in 
Fig. \ref{energyPa} the evolution of ground state energies $E_{0}$ per atom 
along with interaction $U$ and total polarization $P$ in all partially polarized cases. 
The black solid line for $P=0$ here repeats the result of $E_0$ in Fig. \ref{COMenergyp0}.
The fully polarized $P=1$ system is equivalent to the 1D trapped repulsive
scalar bosons in \cite{Hongmei2012} and $E_{0}$ increases monotonously and
approach the ground state energy of non-interacting fermion system.
The energy for partially polarized system with $0<P<1$ interpolates between
these two extremes, i.e., the positive energy terms including 
the kinetic, potential and density-density repulsive interactions together compete with 
the negative binding energy in the anti-ferromagnetically paired bosons, giving 
rise to the non-monotonically dependence of the energy on the interaction parameter 
$U$. In the weak interaction case, the repulsive interaction is prominent so that $E_{0}$ increases
along with $U$. In the strong interaction case, $E_{0}$ decreases because the repulsive
energy slowly increases to a constant whereas the binding energy goes downward parabolically.
Larger polarization destroys the Bose-Bose pairs one by one, which diminishes the effect
of pairing binding energy and finally leads to the monotonic behavior of $E_{0}$ for $P=1$.

\section{Conclusion}

In conclusion, using DFT we study the density distribution and energy of the
1D harmonically trapped spin-1 bosons in the ground state. We numerically
solve the KSEs based on LDA and the solution of Bethe ansatz. The results
show that the competition between the repulsive density-density interaction
and antiferromagnetic spin-exchange interaction results in complicated density 
distributions and energy evolutions along with
the interaction parameter. We found a non-monotonic behavior in the central 
densities of both spin-singlet paired and polarized bosons. Some polarized bosons are 
repelled out of the trap center in the intermediate interaction region, showing 
the double-peak structure of density profiles. The total density exhibits a
bi-modal distribution with paired bosons imposed on the top of polarized bosons. 
The phenomena of phase separation occurs for strong interaction with the
partially polarized core surrounded by wings which are composed of either
paired bosons or polarized bosons depending on the polarization $P$. We give
the R-P phase diagram at strong interaction and find that the
critical polarization $P_{c}$ in DFT is slightly larger than the TFA result. 
Although we treat with an integrable model with equal repulsive
density interaction and antiferromagnetic spin-exchange, the results 
do shed some light on the relativistic spinor gases. We speculate that the new 
quantum phases investigated here could be probed in experiment by 
\textit{in situ} imaging, analogously to the 1D trapped Fermi gas \cite{Liao2010}. 

\begin{acknowledgments}
This work is supported by the NSF of China under Grant Nos. 11234008,
11104171 and 11074153, the National Basic Research Program of China (973
Program) under Grant Nos. 2010CB923103, 2011CB921601, and the Program
for New Century Excellent Talents in University
(NCET). We thank Gao Xianlong, Junpeng Cao and Xiwen Guan for helpful
discussions.
\end{acknowledgments}

\end{document}